\documentstyle[preprint,aps,prb]{revtex}
\tightenlines
\newcommand{\be}{\begin{equation}}
\newcommand{\ee}{\end{equation}}
\newcommand{\ba}{\begin{eqnarray}}
\newcommand{\ea}{\end{eqnarray}}

\begin{document}
\title{Non--Relativistic Spacetimes with Cosmological Constant}

\author{R. Aldrovandi, A. L. Barbosa, L. C. B. Crispino\footnote{Permanent
address: Departamento de F\'{\i}sica,
Universidade Federal do Par\'a, Campus Universit\'ario do Guam\'a,
66075-900 Bel\'em PA, Brazil.} and J. G. Pereira}

\vskip 0.9cm
\address{Instituto de F\'{\i}sica Te\'orica \\
Universidade Estadual Paulista \\
Rua Pamplona 145 \\ 01405-900 S\~ao Paulo \\ Brazil}

\maketitle

\begin{abstract}

Recent data on supernovae favor high values of the
cosmological constant. Spacetimes with a cosmological constant have 
non--relativistic kinematics quite different from  Galilean
kinematics. De Sitter spacetimes, vacuum solutions of Einstein's 
equations with a cosmological constant, reduce in the non--relativistic
limit to Newton--Hooke spacetimes, which are non--metric homogeneous
spacetimes with non--vanishing curvature. The whole non--relativistic
kinematics would then be modified, with possible consequences to cosmology,
and in particular to the missing--mass problem.

\end{abstract}

\vskip 0.8cm

\section{Introduction}

Recent data on supernovae, coming from two independent
programs~\cite{vancou} and favoring high values of the 
cosmological constant, have renewed the interest on the 
non--relativistic limits of the corresponding spacetimes. 
Spacetimes with a cosmological constant have non--relativistic 
kinematics quite different from  Galilean kinematics. They can, 
therefore, modify the non--relativistic physics with further
implications to cosmology, as for example in the use of the virial 
theorem in connection with the missing--mass problem in galaxy clusters. 
Non--relativistic spacetimes have been studied in detail years 
ago.\cite{Kun} We present here another approach to the problem, 
using the technique of group contraction to obtain
non--relativistic kinematics from relativistic  kinematics. 
The coordinate system required by the procedure, which reduces 
to Galilean coordinates in the appropriate
limit, makes this approach nearer to observational practice.

As solutions of the sourceless Einstein's equations with a 
cosmological term, de Sitter spacetimes will play a fundamental 
role. However, as we shall be making good use of
homogeneous spaces in our approach, it will be convenient to start 
our discussion with Lorentzian kinematics. Despite its trivial 
(that is, flat) connection, and as the simplest (flat, vacuum) 
solution of Einstein's equations, Minkowski spacetime $M$ is a true,
even paradigmatic spacetime. It is taken as the local, ``kinematical" 
spacetime and can also be identified with the space tangent to 
(real, curved) spacetime at each point. Above
all, it has a consistent kinematics, in the sense that its metric is 
invariant under the appropriate kinematical (Poincar\'e) group $P$.
This group contains the Lorentz group $L = SO(3,1)$ and includes the 
translation subgroup $T$, which acts transitively on $M$ and is, 
in a sense, its ``double". Indeed, Minkowski spacetime appears as a 
homogeneous space under $P$, actually as the quotient  
$M = T = P/L$. If we prefer, the manifold of $P$ is a principal 
bundle $(P/L, L)$ with $T = M$ as base space and $L$ as the typical 
fiber. 

The invariance of $M$ under the transformations of $P$ reflects its 
uniformity. Also in this ``Copernican" aspect, Minkowski spacetime 
establishes a paradigm.  $P$ has the maximum possible number of Killing 
vectors, which is ten for a 4-dimensional flat spacetime. The Lorentz 
subgroup provides an isotropy around a given point of $M$, and the 
translation invariance enforces this isotropy around any other point. 
This is the meaning of ``uniformity": all the points of spacetime are
ultimately equivalent.

The reduction of relativistic to Galilean kinematics in the 
non--relativistic limit is the standard example of In\"on\"u--Wigner 
contraction,\cite{gilmore} by which the Poincar\'e group is 
contracted to the Galilei group. However, if we insist on
the central role of the metric, there is no such a thing as a real 
``Galilean spacetime". The original metric is somehow ``lost" in 
the process of contraction, and no metric exists which is invariant 
under the Galilei group. Minkowski spacetime tends, in the 
non--relativistic limit, to something that is not a  spacetime. 
Nevertheless, there exists a meaningful connection which survives, even
though the metric becomes undefined. This is not easily visible in the
Minkowski--Galilei case, because both the initial and the final connections 
are flat. 

Actually, in all local, or tangential physics, what happens is that the
laws of Physics are invariant under transformations related to an
uniformity as that described above. It includes homogeneity of space and of
time, isotropy of space and the equivalence of inertial frames. This holds
for Galilean and for special--relativistic physics, their difference being
grounded only in their different ``kinematical groups". However, as was 
clearly shown by Bacry and L\'evy--Leblond,\cite{levy} the corresponding 
Galilei and Poincar\'e groups are not the only ones to answer these uniformity
requirements. Other groups, like the de Sitter groups~\cite{gursey} and their 
non--relativistic In\"on\"u--Wigner contractions, the so called Newton--Hooke 
groups,\cite{dubois} are in principle acceptable candidates.   
	
The complete kinematical group, whatever it 
may be, will always have a subgroup accounting for the isotropy of space
(rotation group) and the equivalence of inertial frames (boosts of
different kinds for each case). The remaining transformations are 
``translations", which may be either commutative or not. Roughly speaking, 
the point--set of the corresponding spacetime is, in each case, the 
point--set of these translations. More precisely, kinematical spacetime 
is defined as the quotient space of the whole kinematical group by the 
subgroup including rotations and boosts. This means that
local spacetime is always a homogeneous manifold. 
	
Amongst curved spacetimes, only those of constant curvature share with
Minkowski spacetime the property of lodging the highest number of Killing
vectors. Given the metric signature and the value of the scalar curvature
$R$, these maximally--symmetric spaces are unique.\cite{weinberg} In
consequence, the de Sitter spacetimes are the only uniformly curved
4-dimensional metric spacetimes. There are two kinds of them,\cite{ellis}
both conformally flat. One of them, the de Sitter
spacetime proper, has the
pseudo--orthogonal group $SO(4,1)$ as group of motions and will be denoted
$DS(4,1)$. The other is the anti--de Sitter spacetime. It will be
denoted $DS(3,2)$ because its group of motions is $SO(3,2)$. They are both
homogeneous spaces: $DS(4,1) = SO(4,1)/ SO(3,1)$ and $DS(3,2) = SO(3,2)/
SO(3,1)$. The manifold of each de Sitter group is a bundle with the
corresponding de Sitter spacetime as base space and $SO(3,1)$ as fiber.

The de Sitter spacetimes are vacuum solutions of Einstein's equations  
with a cosmological constant and, as such, valid alternatives to 
Minkowski spacetime as local kinematical spacetimes.
Provided the de Sitter pseudo--radius parameter $L$ (or inverse cosmological
constant) be large enough, it becomes impossible to 
know whether the true local relativistic group is the Poincar\'e group 
or one of the de Sitter groups, as no experiment could distinguish between 
their consequences. However, if the recent 
cosmological data favoring low values of $L$ comes to be confirmed, 
non--relativistic kinematics would be governed by a  Newton--Hooke  
group, not by the Galilei group.

We shall thus be concerned with such very special kinds of spacetime, the
homogeneous spacetimes of groups which can be called kinematical. Actually, 
we shall be mainly interested in the Newton--Hooke cases, but its study will
require a previous treatment of the Poincar\'e and de Sitter cases,
the Galilei case coming as a corollary. A point we wish to emphasize is that
the general theory of homogeneous spaces warrants the presence of a
connection on such spaces, even when a metric is absent. Applied to our
case, we shall see that the theory endorses the common  knowledge on the 
Minkowski and de Sitter spacetimes, with a metric and corresponding 
connection (with vanishing curvature for Minkowski, but not for de Sitter 
spacetimes). In the Galilei case, it yields no metric and a flat 
connection. In the Newton--Hooke  cases it also gives no metric, but gives
a connection with non--vanishing curvature. 
	
As all cases of our concern will be obtained from the de Sitter cases by
convenient In\"on\"u--Wigner contractions, our main tools
will be the de Sitter groups and spacetimes. For example, the Galilei 
group is obtained from the de Sitter groups by two contractions: one 
which is a non--cosmological  (L $\rightarrow \infty$)  limit, and 
another one which is a non--relativistic  (c $\rightarrow \infty$) limit. 
The Newton--Hooke group, on the other hand, is obtained from the de Sitter 
groups by taking only the non--relativistic limit.
 
Newton--Hooke spacetimes yield the non--relativistic kinematics in 
the case of an eventual non--vanishing cosmological constant, and may 
open the way to some new experimental test allowing to figure out the 
value of $L$. From a more theoretical point of view, the de Sitter 
spacetimes leading to the Newton--Hooke cases can appear in two 
different situations: (i) as spacetime itself, and (ii) as the tangent space 
to spacetime. The latter comes up in gauge theories for the de Sitter 
group, in which the fiber is a de Sitter spacetime tending to the tangent 
space as $L \rightarrow \infty$. Gauge theories for the Poincar\'e 
group~\cite{prd} are, basically, not quantizable, but this difficulty 
is solved if the Poincar\'e group is replaced by a de Sitter
group.\cite{jmp} From this (gauge, quantum) point of view, the latter are
preferable.
	
The theory of homogeneous spaces uses mainly the group Lie algebra, that
is, the algebra of left--invariant fields on the group manifold. The
general theory~\cite{koba} says much more than what is necessary for our
purposes. It contains a whole treatment of invariant connections, and we
only shall need the so called ``canonical" connections. We begin in 
section 2 with a {\it resum\'e} on what the theory of homogeneous spaces
says about those particular connections. To make the formalism easier to
follow, we keep the symbols $``P"$ for the general kinematical group and
$``T"$ for the translational sector, whatever it may be. 
Section 3 is a quick presentation of the conditions for an
invariant metric to exist. We introduce de Sitter spacetimes, and apply 
the algebraic formalism to them, in section 4. 
We find that, for de Sitter spacetimes,
the canonical connection is just the usual, metric connection. 
We find the curvature, but in the Maurer--Cartan basis. 
The tetrad field is then used to obtain the spacetime 
curvature from its corresponding ``algebraic" expression. 
The Newton--Hooke cases are then presented in section 5. It is 
shown that, though no invariant metric exists, a non--trivial 
invariant connection is well--defined, and the non--vanishing 
components of the Riemann tensor are computed.
Some aspects of Newton--Hooke physics are presented in the end of that 
section. Section 6 is dedicated to our final remarks.

\section{The General Algebraic Scheme}

Given a connected Lie group $P$ and a closed subgroup $L$ of $P$, a
homogeneous space is defined by the quotient $P/L$. Consider the Lie
algebras ${\mathcal P}$ of $P$ and ${\mathcal L}$ of $L$. It will be
supposed that a subspace ${\mathcal T}$ of ${\mathcal P}$ exists such that
the underlying vector space of ${\mathcal P}$ is the direct sum ${\mathcal
P} = {\mathcal L} + {\mathcal T}$,
and ${\mathcal T}$ is invariant under the adjoint action of ${\mathcal L}$.
All this means that ${\mathcal P}$ has a multiplication table of the form
\ba
\left[{\mathcal L}, {\mathcal L} \right] &\subset& {\mathcal L} \; ;
\label{21} \\
\left[{\mathcal L}, {\mathcal T} \right] &\subset& {\mathcal T} \; ;
\label{22} \\
\left[{\mathcal T}, {\mathcal T} \right] &\subset& {\mathcal P} =
{\mathcal L} + {\mathcal T} \; .
\label{23}
\ea
This type of Lie algebra, and also the resulting homogeneous space, will
be called ``reductive".\cite{term} This particular case of homogeneous
space includes all spacetimes usually defined from kinematical groups.
Properties (\ref{21}) and (\ref{23}) are essential to an algebraic 
discussion of spacetime, as translations do not constitute a subgroup 
in the general case and $L$ can be seen as a typical fiber. 
	
With our interest in spacetimes in mind, we shall take double--indexed
operators $\{L_{\alpha \beta}\}$ for the generators of $L$ and
simply--indexed $\{T_\gamma\}$ for those of $T$. 
We shall be using the first half of the Greek alphabet
($\alpha, \beta, \gamma, \dots = 0, 1, 2, 3$)
to denote algebraic indices, and the second half
($\lambda, \mu, \nu, \dots = 0, 1, 2, 3$)
to denote spacetime indices.
Thus, the above commutation rules will be
\ba
\left[L_{\gamma \delta}, L_{\epsilon \phi}\right] &=& \frac{1}{2}
f^{(\alpha \beta)}{}_{(\gamma
\delta)(\epsilon \phi)} \, L_{\alpha \beta} \; ; \label{24} \\
\left[L_{\gamma \delta},T_{\epsilon}\right] &=& f^{(\alpha)}{}_{(\gamma
\delta)(\epsilon)} \, T_\alpha \; ; \label{25} \\
\left[T_\gamma, T_\epsilon \right] &=& \frac{1}{2} \, f^{(\alpha
\beta)}{}_{(\gamma)(\epsilon)} \, L_{\alpha \beta} +
f^{(\alpha)}{}_{(\gamma)(\epsilon)} \, T_\alpha
\label{26}
\ea
(the factors $1/2$ account for repeated double--indices). Expression
(\ref{22}) or (\ref{25}) says that the $T$'s belong to a vector
representation of $L$. They are really vectors if they belong to a
commutative algebra. If, as allowed above, they do not, they will be
``extended", that is, they will include a ``connection" with a curvature
along $L$, and torsion along $T$.
	
The canonical form of $P$ will be $\omega = (1/2) L_{\alpha \beta}
\omega^{\alpha \beta} + T_\gamma \omega^\gamma$, where the
$\omega^{\alpha \beta}$'s and $\omega^\gamma$'s are the Maurer--Cartan
forms, dual to the generators. 
The above commutation relations are equivalent to their dual versions, the
Maurer--Cartan equations, which can be stated as: 
\be
{\rm}d\omega + \omega \wedge \omega = 0 \; .
\label{29}
\ee

For us, the most important result of the theory is the following: the
${\mathcal L}$-component of the canonical form of $P$ in that
decomposition defines a connection $\Gamma$ in the bundle $(P/L, L)$
which is invariant under the left--action of $P$:
\be 
\Gamma = \frac{1}{2} (L_{\gamma \delta}) \, \omega^{\gamma \delta} \; .
\label{210}
\ee
The general theory allows many other connections, but this one (called
``canonical") is the most interesting, because of its many fair
properties: \\ 
($i$) it always exists in the reductive case; \\
($ii$) its geodesics are the exponentials of straight lines on the
tangent space; \\ 
($iii$) it is a complete connection; \\
($iv$) its curvature and torsion are parallel-transported; \\
($v$) it transports parallelly any P--invariant tensor. 

Property ($i$) is the most important for our considerations. 
The matrix elements of $\Gamma$ will be 
\be
\Gamma^{\alpha}{}_{\beta} = \frac{1}{2} f^{(\alpha)}{}_{(\gamma
\delta)(\beta)} \, \omega^{\gamma \delta} \; .
\label{211}
\ee
The underlying vector space of
${\mathcal T}$ will be identified with the horizontal space at the 
identity, as it is the set of vector fields $X$ such that 
$\Gamma(X) = 0$. We shall identify the homogeneous 
space to $T$. It should be stressed that (\ref{210}) is a 
Lie--algebra--valued form. The generators $L_{\gamma \delta}$ must
be taken in the representation of interest. On the vector space of the
$T_\alpha$'s, they act as matrices, whose matrix elements are shown in 
(\ref{211}). With this identification, a basis for the forms on $T$ will 
be given by the $\omega^\alpha$'s, and we shall use the notation 
$h = T_\alpha \omega^\alpha$ for the ``horizontal" part of the 
canonical form.

The curvature form $R$ of a connection $\Gamma$ 
is given by
\ba
R = D_{\Gamma}\Gamma = {\rm d}\Gamma + \Gamma \wedge \Gamma \; .
\nonumber
\ea
For the case of reductive homogeneous spacetimes, it 
is fixed by $R(X,Y) Z$ = $- [[X,Y]_{\mathcal L}, Z]$ 
(where $[{~},{~}]_{\mathcal L}$ denotes the ${\mathcal
L}$-component of the commutator) for any left--invariant 
fields  $X,Y,Z \in {\mathcal T}$. Writing the curvature form 
$R$ as
\ba
R =  \frac{1}{4} \, L_{\alpha \beta} \, R^{\alpha
\beta}{}_{\gamma \delta} \,
\omega^{\gamma} \wedge \omega^{\delta} \; ,
\nonumber
\ea
we obtain for its components
\be
R^{\alpha \beta}{}_{\gamma \delta} = - \, f^{(\alpha 
\beta)}{}_{(\gamma) (\delta)} \; .
\label{213}
\ee

The torsion of a connection $\Gamma$ is given by
the covariant derivative of the basis $h$ according to $\Gamma$:
\ba 
\Theta = D_{\Gamma}h = {\rm d}h + \Gamma \wedge h + h \wedge \Gamma \, .
\nonumber 
\ea   
For reductive homogeneous spacetimes, the
torsion is fixed by $\Theta(X,Y) = - [X,Y]_{\mathcal T}$, so
that a non--vanishing torsion requires that some
$f^{(\alpha)}{}_{(\gamma)(\delta)} \neq 0$. 
Writing the torsion form $\Theta$ as 
\ba
\Theta = \frac{1}{2} \, T_{\alpha} \, \Theta^{\alpha}{}_{\gamma \delta} \, 
\omega^{\gamma} \wedge \omega^{\delta} \; ,
\nonumber
\ea
we obtain for its components
\be
\Theta^{\alpha}{}_{\gamma \delta} = - f^{(\alpha)}{}_{(\gamma)(\delta)} \; .
\label{215}
\ee
Homogeneous spacetimes with torsion can appear when invariance under
parity and time reversal transformations are ignored.\cite{bacry}

It should be said that, conversely, any connection on the bundle
$(P/L,L)$, which is invariant under the left--action of $P$, determines a
decomposition as above.
	
The general properties can therefore be obtained by inspecting the
multiplication table.
One drawback of the theory is that the connection and its curvature and 
torsion come out naturally in a particular basis, that constituted by 
the Maurer--Cartan forms of the whole group.

\section{Invariant Metrics}

The above connection is invariant under the action of $P$.  
However, in the general case, it has nothing to do with a
metric. The scheme does provide for metrics, and establishes conditions
under which the connection is metric. Actually there will be a
($P$-invariant) metric on $T$ for each ${\rm ad}_{L}$-invariant
non--degenerate symmetric bilinear form $B$ on $P/L: g(X,Y) = B(X,Y)$ for
all $X,Y \in {\mathcal T}$. The required invariance under
${\rm ad}_{L}$ is expressed by 
\ba
B \left([Z,X],Y\right) +  B \left(X,[Z,Y]\right) = 0 \; ,
\nonumber
\ea
for all $X,Y \in {\mathcal T}$ and $Z \in {\mathcal L}$. Thus, with
$B_{\alpha \beta} = B(T_\alpha, T_\beta)$,
\be
B_{\delta \epsilon} \, f^{(\epsilon)}{}_{(\alpha \beta)(\gamma)} +
B_{\gamma
\epsilon} \, f^{(\epsilon)}{}_{(\alpha \beta)(\delta)} = 0 \; .
\label{32}
\ee
{}From (\ref{211}), this is the same as 
\be
B_{\delta \epsilon} \,
\Gamma^{\epsilon}{}_{\gamma} + B_{\gamma \epsilon} \,
\Gamma^{\epsilon}{}_{\delta} = 0 \; . 
\label{con}
\ee
If $B$ is a metric, so that we can
lower indices, it becomes 
$\Gamma_{\delta \gamma} = - \, \Gamma_{\gamma \delta} \,$ .

This is typical of (pseudo-)orthogonal connections, 1-forms with values
in the Lie algebra of (pseudo-)orthogonal groups. Each bilinear form
satisfying this condition gives an invariant metric. This is clearly the
case for the de Sitter groups and for $P$ a semisimple group in general:
the Cartan--Killing metric will do. We know, on the other hand, that the
Killing form is degenerate for non--semisimple groups. However, any other
such bilinear can lead to a metric. It will be shown below how this
happens for the Poincar\'e group case, for which the Lorentz metric will
come up naturally.

When we say, rather loosely, that a metric exists, or not, we mean that 
there exists, or not, a metric which is invariant under the kinematical 
group of transformations. Thus, we shall see that there is no metric on 
Galilean ``spacetime" which is invariant under the Galilei group, and 
no metric on the Newton--Hooke spacetime which is invariant under the 
Newton--Hooke transformations. The same invariance requirement holds 
for connections. In both cases above there will be invariant connections 
(though trivial in the Galilean case).

\section{The de Sitter Cases} 
	
We start by introducing the de Sitter spacetime $DS(4,1)$ and the (anti-) 
de Sitter spacetime $DS(3,2)$ as hypersurfaces in the pseudo--Euclidean 
spacetimes
${\bf E}^{4,1}$ and
${\bf E}^{3,2}$, inclusions whose points in Cartesian coordinates
$(\xi) = (\xi^0, \xi^1, \xi^2, \xi^3$, $\xi^4)$ satisfy, respectively,
\ba
(\xi^1)^2 + (\xi^2)^2 + (\xi^3)^2 - (\xi^0)^2 +
(\xi^4)^2 &=& L^2 \; ; \nonumber \\
(\xi^1)^2 + (\xi^2)^2 + (\xi^3)^2 - (\xi^0)^2 -
(\xi^4)^2 &=& - L^2 \; . \nonumber
\ea
We use $\eta_{\alpha \beta}$ for
the Lorentz metric $\eta = {\rm diag} (-1, 1, 1, 1)$ and the notation
$\eta_{44} = \epsilon$ to put the conditions together as
\[
\eta_{\alpha \beta} \, \xi^{\alpha} \xi^{\beta} + \epsilon
\left(\xi^4\right)^2 =
\epsilon L^2 \; .
\]
We now change to stereographic coordinates $\{x^\mu\}$ in 4--dimensional
space, which are given by
\be
x^\mu = h_{\alpha}{}^{\mu} \, \xi^{\alpha} \; ,
\label{41}
\ee
where
\be
h^{\alpha}{}_{\mu} = n \, \delta^{\alpha}{}_{\mu} \; ;
\label{42}
\ee
\be
n = \frac{1}{2} \, \left(1 - \frac{\xi^4}{L} \right) = \frac{1}{1+
\epsilon \sigma^2 / 4 L^2} \; ,
\label{43}
\ee
with $\sigma^2 = \eta_{\alpha \beta} \, \delta^{\alpha}{}_{\mu}
\delta^{\beta}{}_{\nu} \, x^\mu x^\nu$. Calculating the line element
for the de Sitter spacetimes, we find $ds^2 = g_{\mu \nu} \,d x^\mu
dx^\nu$, where
\be
g_{\mu \nu} = h^{\alpha}{}_{\mu} h^{\beta}{}_{\nu} \eta_{\alpha \beta} \;
.
\label{44}
\ee
The $h^{\alpha}{}_{\mu}$ given in (\ref{42}) are the components of the
tetrad field, actually of the 1-form basis members
\be
\omega^{\alpha} = h^{\alpha}{}_{\mu} dx^\mu = 
n \delta^{\alpha}{}_{\mu} dx^\mu \; ,
\ee
dual to a vector basis
\be
e_{\alpha} = h_{\alpha}{}^{\mu} \partial_\mu = 
\frac{1}{n} \delta_{\alpha}{}^{\mu} \partial_\mu \; .
\label{duba}
\ee 
The de Sitter spacetimes are conformally flat, with the conformal 
factor $n^2(x)$ given by (\ref{43}). We could have written simply 
$x^\mu = \xi^\mu / n$, but we are carefully using the first letters 
of the Greek alphabet for the algebra (and flat space) indices, and 
those from the middle on for the homogeneous space fields and
cofields. A certain fastidiousness concerning indices is necessary 
to avoid any confusion in the later contraction to the Newton--Hooke 
cases, and accounts for some apparently irrelevant $\delta$'s in some 
expressions. As usual with changes from flat tangent--space to spacetime, 
letters of the two kinds are interchanged with
the help of the tetrad field: for example,
$f^{(\sigma)}{}_{(\alpha \beta)(\mu)} = h_{\gamma}{}^{\sigma}
f^{(\gamma)}{}_{(\alpha \beta)(\epsilon)} h^{\epsilon}{}_{\mu}$. This is
true for all tensor indices. Connections, which are vectors only in the
last (1-form) index, acquire an extra ``vacuum" term:
\be
\Gamma^{\lambda}{}_{\mu \nu} = h_{\alpha}{}^{\lambda} \,
\Gamma^{\alpha}{}_{\beta
\nu} \, h^{\beta}{}_{\mu} + h_{\gamma}{}^{\lambda} \partial_\nu
h^{\gamma}{}_{\mu} \; .
\label{45}
\ee

The Christoffel symbols corresponding to the metric $g_{\mu \nu}$ are
\be
\Gamma^{\lambda}{}_{\mu \nu} = \left[ \delta^{\lambda}{}_{\mu}
\delta^{\sigma}{}_{\nu}  + \delta^{\lambda}{}_{\nu}
\delta^{\sigma}{}_{\mu} - g_{\mu \nu} g^{\lambda \sigma} \right]
\partial_\sigma (\ln n) \; .
\label{46}
\ee
The Riemann tensor components, $R^{\alpha}{}_{\beta \rho \sigma} =
\partial_\rho
\Gamma^{\alpha}{}_{\beta \sigma} - \partial_\sigma
\Gamma^{\alpha}{}_{\beta \rho} +
\Gamma^{\alpha}{}_{\epsilon \rho} \, \Gamma^{\epsilon}{}_{\beta \sigma} -
\Gamma^{\alpha}{}_{\epsilon \sigma} \, \Gamma^{\epsilon}{}_{\beta \rho}$,
are found to be
\be
R^{\alpha}{}_{\beta \rho \sigma} = \epsilon \frac{1}{L^2}
\delta_{\beta}{}^{\mu}
\left[\delta^{\alpha}{}_{\rho} g_{\mu \sigma} - 
\delta^{\alpha}{}_{\sigma} g_{\mu \rho} \right] \; .
\label{47}
\ee
The Ricci tensor and the scalar curvature are, consequently
\be
R_{\mu \nu} = \epsilon \frac{3}{L^2} g_{\mu \nu} \; ;
\label{48}
\ee
\be
R = \epsilon \frac{12}{L^2} \;.
\label{49}
\ee
Minkowski quantities are found in the contraction limit $L \rightarrow
\infty$. Galilean kinematics comes then from the subsequent contraction $c
\rightarrow \infty$. As usual with contractions, some infinities are
absorbed in new, redefined, parameters. The Newton--Hooke cases are attained
by taking the non--relativistic $c \rightarrow \infty$ limit while keeping
finite an appropriate time parameter $\tau = L/c$ (see section V).
	
Let us now examine the Lie algebra of the de Sitter groups. The Lorentz
sector is given by
\be
[L_{\alpha \beta},L_{\gamma \delta}] = \frac{1}{2} \, f^{(\epsilon
\phi)}{}_{(\alpha
\beta)(\gamma \delta)} \, L_{\epsilon \phi} \; ,
\label{410}
\ee
with
\be
f^{(\epsilon \phi)}{}_{(\alpha \beta)(\gamma \delta)} = \eta_{\beta \gamma}
\delta^{\epsilon}{}_{\alpha} \delta^{\phi}{}_{\delta} + \eta_{\alpha \delta}
\delta^{\epsilon}{}_{\beta} \delta^{\phi}{}_{\gamma} - \eta_{\beta \delta}
\delta^{\epsilon}{}_{\alpha} \delta^{\phi}{}_{\gamma} - \eta_{\alpha \gamma}
\delta^{\epsilon}{}_{\beta} \delta^{\phi}{}_{\delta} \; .
\label{411}
\ee
The de Sitter translation generators are Lorentz vectors,
\be
[L_{\alpha \beta},T_{\gamma}] = f^{(\epsilon)}{}_{(\alpha \beta)(\gamma)} \,
T_\epsilon \; ,
\label{412}
\ee
with
\be
f^{(\epsilon)}{}_{(\alpha \beta)(\gamma)} = 
\eta_{\gamma \beta} \delta^{\epsilon}{}_{\alpha} - 
\eta_{\gamma \alpha} \delta^{\epsilon}{}_{\beta} \; .
\label{413}
\ee
Up to this point, of course, the situation is identical to that of the
Poincar\'e group. For the invariant metric also, as only the values of
$f^{(\epsilon)}{}_{(\alpha \beta)(\gamma)}$ are involved, the same bilinear
will work for both de Sitter and Poincar\'e. For the translation sector,
we shall have
\be
[T_\alpha,T_\beta] = \frac{1}{2} f^{(\epsilon \phi)}{}_{(\alpha)(\beta)} \,
L_{\epsilon \phi} + f^{(\gamma)}{}_{(\alpha)(\beta)} \, T_\gamma \; ,
\label{414}
\ee
with
\be
f^{(\epsilon \phi)}{}_{(\alpha)(\beta)} = - \frac{\epsilon}{L^2}
\left( \delta^{\epsilon}{}_{\alpha} \delta^{\phi}{}_{\beta} -
\delta^{\phi}{}_{\alpha} \delta^{\epsilon}{}_{\beta} \right) \; ;
\label{415}
\ee
\be
f^{(\gamma)}{}_{(\alpha)(\beta)} = 0 \; .
\label{416}
\ee
The contraction factor $1/L^2$ has already been introduced in (\ref{415}). 
In the contraction limit $L \rightarrow \infty$, it gives the usual 
commutative translations of the Poincar\'e group. The vanishing in 
(\ref{416}) accounts for the absence of torsion in all cases we shall 
be concerned with.
	
Concerning the existence of an invariant metric, the condition (\ref{32})
will be
\be 
\eta_{\gamma \beta} B_{\delta \alpha} - \eta_{\gamma \alpha} B_{\delta
\beta} +
\eta_{\delta \beta} B_{\gamma \alpha} - \eta_{\delta \alpha} B_{\gamma
\beta} = 0 \; .
\label{418}
\ee
This is satisfied, in particular, by $B_{\alpha \beta} = \eta_{\alpha
\beta}$, which is quite natural: the group is usually introduced by this
condition. The usual condition defining an orthogonal group says that $B A
B^{-1} = - A^T$, for $A$ any member of the group Lie algebra. Take the
generators $J_{\gamma \delta}$: the condition becomes $B_{\alpha \beta}
(J_{\gamma 
\delta})^{\beta}{}_{\epsilon} B^{\epsilon \phi} = - (J_{\gamma 
\delta})^{\phi}{}_{\alpha}$ which implies
$B_{\alpha \beta}f^{(\beta)}{}_{(\gamma \delta)(\epsilon)} B^{\epsilon
\phi} = - f^{(\phi)}{}_{(\gamma \delta)(\alpha)}$. Lowering and raising
indices with
$B$ and $B^{-1}$, we find $f_{(\alpha)(\gamma \delta)}{}^{(\phi)} = -
f^{(\phi)}{}_{(\gamma \delta)(\alpha)}$, equivalent to (\ref{32}). Thus,
for (pseudo-)orthogonal groups, $B$ can always be the original preserved
metric. In the present case, (\ref{32}) is also satisfied by any metric 
of the form $B_{\alpha \beta} = n^2 \eta_{\alpha \beta}$, with $n^2$ 
positive.

Using the fact that the metric $\eta_{\alpha \beta}$ is invariant
under the action of the de Sitter group, we can use (\ref{con})
and the transformation rule (\ref{45}) to write
\ba
D_{\lambda} g_{\mu \nu} =
\partial_{\lambda} g_{\mu \nu}
- \Gamma^{\sigma}{}_{\mu \lambda} g_{\sigma \nu} 
- \Gamma^{\sigma}{}_{\nu \lambda} g_{\mu \sigma} = 0 \; ,
\nonumber
\ea
which says that the de Sitter canonical connection 
$\Gamma^{\lambda}{}_{\mu \nu}$, with torsion
\[ 
\Theta = - \frac{1}{2} \, T_\alpha \, 
f^{(\alpha)}{}_{(\gamma)(\delta)} \, \omega^{\gamma} \wedge
\omega^{\delta} = 0 \, ,
\]
parallel-transports the de Sitter
spacetime metric $g_{\mu \nu}$ given in (\ref{44}). According to
Ricci's theorem~\cite{ap} there is a unique linear 
connection $\Gamma$ which preserves a metric $g$ and has a fixed
$\Theta$ for its torsion. Therefore, we conclude that the canonical 
connection of the de Sitter spacetimes regarded as reductive
homogeneous spacetimes is exactly the same as the de Sitter
Christoffel symbols (\ref{46}) obtained by requiring {\it a priori}
the connection to be torsionless and the covariant derivative
of the metric (\ref{44}) to be zero.
Starting from (\ref{46}), and using the transformation rule
(\ref{45}), we find
\be
\Gamma_{\alpha \beta \gamma} = \frac{1}{n} \, \left[\eta_{\alpha \gamma}
\delta^{\epsilon}{}_{\beta} - \eta_{\beta \gamma}
\delta^{\epsilon}{}_{\alpha} \right] e_{\epsilon}(n) \; ,
\label{dSac}
\ee
with $e_{\epsilon}(n)$ the vector basis (\ref{duba}) applied to $n$. 
	
This canonical connection has algebraic curvature with com\-po\-nents 
$R^{\alpha \beta}{}_{\gamma \delta}$ = $- \, f^{(\alpha 
\beta)}{}_{(\gamma) (\delta)}$, from which we obtain
\be
R^{\alpha}{}_{\beta \gamma \delta} = 
\eta_{\beta \phi} R^{\alpha \phi}{}_{\gamma \delta} =
\frac{\epsilon}{L^2} \, \left[\eta_{\beta
\delta} \delta^{\alpha}{}_{\gamma} - \eta_{\beta \gamma}
\delta^{\alpha}{}_{\delta}
\right] \; .
\label{419}
\ee
The algebraic Ricci tensor will be
\be
R_{\alpha \beta} = \epsilon \frac{3}{L^2} \eta_{\alpha \beta} \; .
\label{420}
\ee
By using the tetrad field, the components of
the geometrical curvature can be obtained by
\ba
R^{\alpha}{}_{\beta \mu \nu} = 
R^{\alpha}{}_{\beta \gamma \delta} \, 
h_{\gamma}{}^{\mu} \,
h_{\delta}{}^{\nu} \; .
\nonumber
\ea
We thus obtain (\ref{47}), and subsequently (\ref{48}) and (\ref{49}). 

It has been discussed in the literature whether the connection determines 
the metric and whether curvature determines the connection.\cite{arb}
In the case of the de Sitter spacetimes we can use the homotopy
formula~\cite{ap} to obtain the connection from their Riemann tensor, 
and the metric from their connection. As this discussion is out of the 
scope of the present paper, we shall not present it here.

\section{The Newton--Hooke Spacetimes}

Newton--Hooke spacetimes can be considered as the non--relativistic limits
of the de Sitter ones. Their main characteristic is that the spacetime 
translations contain a global effect inherited from the de Sitter
curvature. This is due to the fact that, in contrast to the Galilei group,
no non--cosmological limit is taken to get the Newton--Hooke groups.

To obtain the Newton--Hooke algebras, we submit de Sitter multiplication
table to a In\"on\"u--Wigner contraction. In
the contraction of a Lie algebra, we must ensure the desired behavior of
the limiting generators through a careful choice of parameterization, which
reflects itself in the structure coefficients. This is better understood
if we take a particular representation. Let us consider the so--called
kinematical representation, in which the de Sitter generators are given by
vector fields on ${\bf E}^{3,2}$ and ${\bf E}^{4,1}$:
\be
L_{AB} = \xi_A \frac{\partial}{\partial \xi^B}  - \ \xi_B
\frac{\partial}{\partial \xi^A} \; ,
\label{51}
\ee
with the indices $A,B = 0,1,2,3,4$ .
We then pass to the stereographic coordinates (\ref{41}), with the
identifications
\ba
(x^i) &=& (x, y, z) : {\rm Cartesian \, coordinates \, in} \, {\bf E}^3 \; ; 
\nonumber \\
x^0 &=& c t \; . \nonumber
\ea
{}From now on, we shall be using $a, b, c, \dots = 1, 2, 3$ for the algebra
indices, and $i, j, k,\dots = 1, 2, 3$ for the space indices. The
non--relativistic case is achieved in the limit $c \rightarrow \infty$,
after appropriate redefinitions of quantities which would otherwise
exhibit divergences. Here, we must first separate time and space
components of $L_{AB}$, obtaining explicit forms for $L_{ab}$,
$L_{a0}$, $L_{a4}$ and $L_{04}$, and then redefine these operators so as
to have finite expressions in the limit. In the present case, we must go
through the intermediate step of introducing a new time parameter $\tau =
L/c$ which remains finite in the process. The redefinitions are the
following:
\be
{\mathsf L}_{ab} \equiv L_{ab} \; , \; {\mathsf L}_{a0} \equiv L_{a0}/c 
\; , \;
{\mathsf T}_a \equiv \epsilon L_{a4}/c \tau \; , \; {\mathsf T}_0 \equiv
\epsilon  L_{04}/\tau \; .
\label{52}
\ee
This corresponds to introducing the inverse factors in the parameters, so
that $\omega^{ab} \rightarrow \omega^{ab}$; $\omega^{a0} \rightarrow c
\omega^{a0}$; $\omega^{a} \rightarrow \epsilon c \tau \omega^{a}$, and
$\omega^{0} \rightarrow \epsilon \tau \omega^{0}$. The factors are then
absorbed in the redefined parameters, which acquire different dimensions.
Connections behave like parameters (more precisely: like the dual
Maurer--Cartan forms, which behave like parameters), so that a connection
component $\Gamma^{a 0}$, for example, will acquire a factor
$(1/c)$. The de Sitter conformal function (\ref{43}) becomes
\be
n(t) = \frac{1}{1 - \epsilon t^2/4 \tau^2} \; ,
\label{53}
\ee
and the tetrad fields will consequently be
\be
h^{\alpha}{}_{\mu} = \frac{\delta^{\alpha}{}_{\mu}}{1 - \epsilon t^2/4
\tau^2} \; .
\label{54}
\ee

In terms of the redefined generators, the de Sitter multiplication table
(\ref{410})-(\ref{416}) becomes
\be
\left[{\mathsf L}_{ab}, {\mathsf L}_{de}\right] = \delta_{bd} {\mathsf
L}_{ae} + \delta_{ae} {\mathsf L}_{bd} - \delta_{be} {\mathsf L}_{ad} -
\delta_{ad} {\mathsf L}_{be} \; ;
\label{55}
\ee
\be
\left[{\mathsf L}_{ab}, {\mathsf L}_{d0}\right] = \delta_{bd} {\mathsf
L}_{a0} - \delta_{ad} {\mathsf L}_{b0} \; ;
\ee
\be
\left[{\mathsf L}_{0b}, {\mathsf L}_{0e}\right] = \frac{1}{c^2}  {\mathsf
L}_{be} \; ;
\ee
\be
\left[{\mathsf L}_{ab},{\mathsf T}_{d}\right] = \delta_{bd} {\mathsf
T}_{a} -
\delta_{ad} {\mathsf T}_{b} \; ;
\ee
\be
\left[{\mathsf L}_{a0}, {\mathsf T}_{b}\right] =  \frac{1}{c^2}
\delta_{ab} {\mathsf T}_{0} \; ;
\ee
\be
\left[{\mathsf L}_{a0}, {\mathsf T}_{0}\right] = - {\mathsf T}_{a} \; ;
\ee
\be
\left[{\mathsf L}_{ab}, {\mathsf T}_{0}\right] = 0 \; ;
\ee
\be
\left[{\mathsf T}_a, {\mathsf T}_b\right] = - \frac{\epsilon}{\tau^2 c^2}
{\mathsf L}_{ab} \; ;
\ee
\be
\left[{\mathsf T}_a, {\mathsf T}_0\right] =  - \frac{\epsilon}{\tau^2} 
{\mathsf L}_{a0} \; ;
\label{TaT0DS}
\ee
\be
\left[{\mathsf T}_0, {\mathsf T}_0\right] = 0 \; .
\ee
This is the appropriate parameterization, in which the Newton--Hooke
algebra is obtained by taking the limit $c \rightarrow \infty$. The result
is then
\be
\left[{\mathsf L}_{ab}, {\mathsf L}_{de}\right] = \delta_{bd} {\mathsf
L}_{ae} +
\delta_{ae} {\mathsf L}_{bd} - \delta_{be} {\mathsf L}_{ad} - \delta_{ad}
{\mathsf L}_{be} \; ;
\ee
\be
\left[{\mathsf L}_{ab}, {\mathsf L}_{d0}\right] = \delta_{bd} {\mathsf
L}_{a0} -
\delta_{ad} {\mathsf L}_{b0} \; ;
\ee
\be
\left[{\mathsf L}_{0b}, {\mathsf L}_{0e}\right] = 0 \; ;
\ee
\be
\left[{\mathsf L}_{ab},{\mathsf T}_{d}\right] = \delta_{bd} {\mathsf
T}_{a} -
\delta_{ad} {\mathsf T}_{b} \; ;
\ee
\be
\left[{\mathsf L}_{a0}, {\mathsf T}_{b}\right] = 0 \; ;
\ee
\be
\left[{\mathsf L}_{a0}, {\mathsf T}_{0}\right] = - {\mathsf T}_{a} \; ;
\ee
\be
\left[{\mathsf L}_{ab}, {\mathsf T}_{0}\right] = 0 \; ;
\ee
\be
\left[{\mathsf T}_a, {\mathsf T}_b\right] = 0 \; ;
\ee
\be
\left[{\mathsf T}_a, {\mathsf T}_0\right] = - \frac{\epsilon}{\tau^2}
{\mathsf L}_{a0} \; ;
\label{523}
\ee
\be
\left[{\mathsf T}_0, {\mathsf T}_0\right] = 0 \; .
\ee
Vanishing torsion is inherited from the de Sitter cases. 
We see from (\ref{523}) an important difference with respect to the 
Galilean case: time translations do not commute with space translations. 
In consequence, there will be constants surviving contraction in the 
curvature (\ref{213}). Such non--vanishing components of the
algebraic curvature $R^{\alpha \beta}{}_{\gamma \delta}$
will be of the form
\be
R^{a0}{}_{b0} = \frac{\epsilon}{\tau^2} \delta^a{}_{b} \; ,
\label{524}
\ee
and those obtained by antisymmetrizing in the index--pairs. 
We obtain the non--vanishing geometrical curvature components
$R^{\alpha \beta}{}_{\mu \nu}$ from (\ref{524}) by using the tetrad
field (\ref{54}):
\ba 
R^{a0}{}_{i0} = - \, R^{a0}{}_{0i} = 
- \, R^{0a}{}_{i0} = R^{0a}{}_{0i} = 
\frac{\epsilon}{\tau^2} \delta^{a}{}_{i} \, n^2(t) \; .
\nonumber
\ea
Thus, the Newton--Hooke ``spacetimes" are curved, in the sense that their
canonical connections as homogeneous spaces have non--vanishing curvature
Riemann tensor.

To show the absence of metric
we notice that, of all the constants
appearing in condition (\ref{32}), the only ones surviving contraction are
$f^{(a)}{}_{(de)(b)} = (\delta_{eb} \delta_{d}{}^{a} - \delta_{db}
\delta_{e}{}^{a})$ and $f^{(a)}{}_{(b0)(0)} = - \delta^a{}_{b}$. Then,
from
\[
B_{\delta a} f^{(a)}{}_{(\alpha \beta)(\gamma)} + B_{\gamma a}
f^{(a)}{}_{(\alpha
\beta)(\delta)} = 0 \; ,
\]
a component--by--component analysis shows that $B$ must have $B_{ab} = 0$
and $B_{0b} = 0$. There is no condition on the component $B_{00}$, which
can assume any value. Anyhow, the bilinear form will be degenerate. No 
metric is possible. There is no spacetime in the usual, metric sense. 
	
The procedure extends immediately to the Galilei case. 
Galilean spacetime is, however, flat in the same sense in
which the Newton--Hooke spacetime is curved: the Galilean canonical
connection has vanishing Riemann tensor.
Notice that to have a flat connection is quite different from having
no connection at all. Galilean connection is flat, but it exists.

For completeness we have calculated explicitly the canonical connection of
Newton--Hooke spacetimes. Taking the appropriate limit of 
the de Sitter canonical connection (\ref{dSac}),
we obtain the non--vanishing components of the Newton--Hooke connection
$\Gamma^{\alpha \beta}(x)$, namely
\ba
\Gamma^{a0}(x) = - \, \Gamma^{0a}(x) = 
- \, \frac{\epsilon}{2 \tau^2} \, \delta^{a}{}_{i} \left[t
\, dx^i - x^i \, dt \right] n(t) \; .
\label{65}
\ea

The Newton--Hooke spacetimes appear then as examples of non--metric
curved spacetimes. Despite the non--relativistic limit, the effects of 
curvature are present due to the fact that we are still considering a 
kinematical spacetime on a large scale of time.
Notwithstanding, in such a spacetime there is an absolute time.
As in the Galilean case, simultaneity of two events is preserved 
by a general inertial transformation.
Another interesting physical feature of Newton--Hooke spacetimes is that,
analogously to the de Sitter cases, energy is not invariant under
spatial translations. This can be directly verified from
Eqs.~(\ref{TaT0DS}) and (\ref{523}).

The explicit form of the infinitesimal generators for rotations
and boosts are the same for Newton--Hooke and Galilei groups. For the
spacetime translations generators, Newton--Hooke differ from Galilei  
by factors proportional to $\tau ^{-2}$, which vanish in the 
$\tau \rightarrow \infty$ limit.

\section{Final Remarks}
	
Newton--Hooke groups and spacetimes appear as the non--relativistic limits
in all theories involving de Sitter groups and spacetimes, in the same way
the Galilei group appears as the corresponding non--cosmological
non--relativistic limit. Therefore, if the relativistic spacetime 
kinematical group were supposed to be de Sitter instead of Poincar\'e, 
the non--relativistic limit would lead to Newton--Hooke instead of Galilei, 
the latter being obtained through
a further non--cosmological contraction. It is commonly accepted that
non--relativistic physics is invariant under the Galilei group, and it 
is known since
long that the cosmological constant is very small inside the solar system.
Discrepancies between values of the Hubble constant obtained from nearby and
distant objects have led, however, to the proposal that the ``local" values
(cosmologists say ``small scale" values) of cosmological quantities are not
typical, and could be different in other regions of the 
Universe.\cite{BBST95} Newton--Hooke spacetimes are, thus, fair candidates 
for non--relativistic physical
spacetimes in regions in which the value of $\tau$ is large. 

The main differences of Newton--Hooke with respect to Galilei kinematics 
come from the non--commutativity between space and time translations. 
Usual non--relativistic time--evolution equations, for example, in 
which the 3-space Laplacian is time--invariant, would be changed. Local 
Newton--Hooke spacetimes can, therefore, lead to important modifications 
in cosmological non--relativistic physics, as used in
the approach to galaxy clusters via the virial theorem, or in the study
of the rotation spectra of individual galaxies.

There are further, more conceptual aspects.  In its standard definition, 
a spacetime is a 4-manifold with a Lorentzian metric $g$. More precisely, 
it is a pair $({\mathcal M},g)$ where ${\mathcal M}$ is a connected 
4-dimensional differentiable manifold and $g$ is a metric of signature 
2 on ${\mathcal M}$. The metric provides 
a natural torsionless Levi--Civita connection, with a Riemann curvature 
associated to it, but the fact remains that curvature is a connection, 
metric--independent, characteristic. The existence of curved spacetimes 
on which no metric is defined suggests that ``spacetime" should be defined  
not as a pair $({\mathcal M},g)$, a manifold plus a metric,
but as a pair $({\mathcal M},\Gamma)$ --- a differentiable manifold plus 
a connection.
Galilean spacetime would in that case acquire a well--defined status, 
despite its flat connection. Gravitation would be brought closer to the 
other fundamental interactions of Nature, all of them described by 
connections. In the standard cases of metric spacetimes, the connection 
would be the Christoffel symbols obtained by
requiring it to parallel--transport the metric and to have zero torsion. 
A connection satisfying these two conditions 
is unique, according to Ricci's theorem. 
Consequently, when spacetime is also a homogeneous space, 
this connection coincides with the torsionless canonical connection 
obtained following the general algebraic scheme of
section 2, which complies with both of the above conditions.

\section*{Acknowledgements}

One of us (LCBC) has been supported by CAPES (Bras\'{\i}lia), through a
PICDT program. The others (RA, ALB, JGP) have been supported by CNPq
(Bras\'{\i}lia). The authors are grateful to S. F. Novaes for helpful 
discussions on the supernovae data.

\end{document}